\documentclass[aps,english,showpacs]{revtex4}%
\usepackage[T1]{fontenc}
\usepackage{amsmath}
\usepackage{graphicx}
\usepackage{amsfonts}
\usepackage{amssymb}%
\providecommand{\U}[1]{\protect\rule{.1in}{.1in}}
\providecommand{\U}[1]{\protect\rule{.1in}{.1in}}
\providecommand{\U}[1]{\protect\rule{.1in}{.1in}}
\providecommand{\U}[1]{\protect\rule{.1in}{.1in}}
\makeatletter
\makeatletter
\providecommand{\U}[1]{\protect\rule{.1in}{.1in}}
\makeatother
\makeatother
\begin{document}
\title{Duality for massive spin two theories in arbitrary dimensions }
\author{B. Gonz\'alez$^{1}$, A. Khoudeir$^{2}$, R. Montemayor$^{3}$ and L. F.
Urrutia$^{1}$}
\affiliation{$^{1}$Instituto de Ciencias Nucleares, Universidad Nacional Aut\'onoma de
M\'exico, A. Postal 70-543, 04510 M\'exico D.F.}
\affiliation{$^{2}$Centro de F\'\i sica Fundamental, Departamento de F\'\i sica, Facultad
de Ciencias, Universidad de Los Andes, M\'erida 5101, Venezuela}
\affiliation{$^{3}$Instituto Balseiro and CAB, Universidad Nacional de Cuyo and CNEA, 8400
Bariloche, Argentina}

\begin{abstract}
Using the parent Lagrangian approach we construct a dual formulation, in the
sense originally proposed by Curtright and Freund, of a massive spin two
Fierz-Pauli theory in arbitrary dimensions $D$. This is achieved in terms of a
mixed symmetry tensor $T_{A[B_{1}B_{2}\dots B_{D-2}]}$, without the need of
auxiliary fields. The relation of this method with an alternative formulation
based on a gauge symmetry principle proposed by Zinoviev is elucidated. We
show that the latter formulation in four dimensions, with a given gauge fixing
together with a definite sequence of auxiliary fields elimination via their
equations of motion, leads to the parent Lagrangian already considered by West
completed by a Fierz-Pauli mass term, which in turns yields the
Curtright-Freund action. This motivates our generalization to arbitrary
dimensions leading to the corresponding extension of the four dimensional
result. We identify the transverse true degrees of freedom of the dual theory
and verify that their number is in accordance with those of the massive
Fierz-Pauli field.

\end{abstract}

\maketitle

\section{Introduction}

Fields with spin two and higher in dimensions larger than four are of
considerable interest for understanding supersymmetric string theories
together with their brane extensions from the perspective of the M-theory. An
additional feature that adds interest to this problem is that in dimensions
$D>5$, the totally symmetric tensor fields are not enough to cover all the
irreducible representations of the Poincar\'{e} group. Thus, when dealing with
higher spin theories it becomes necessary to take into account fields with
mixed symmetry \cite{CURT1,CURT,GEN} belonging to \textquotedblleft
exotic\textquotedblright\ representations of the Poincar\'{e} group. It is
therefore quite natural to expect that, in a field theory limit, the
superstring theory should reduce to a consistent interacting supersymmetric
theory of massless and massive higher spin fields. In four space-time
dimensions, Lagrangian formulations for massive fields of arbitrary spin were
originally constructed in Ref. \cite{Singh}. Later, this construction was used
to derive Lagrangian formulations for massless gauge fields of arbitrary spin
\cite{Fronsdal}. An important matter related to mixed symmetry tensor fields
is the study of their consistent interactions, among themselves as well as
with higher-spin gauge theories \cite{Bengtsson}. Amid the many approaches to
the problem, a particularly interesting one is the Zinoviev approach
\cite{ZINOVIEV1} where the gauge symmetry principle has been extended to deal
with the massive case in a way that incorporates a Stueckelberg-like
formulation of the corresponding actions in the background of Minkowski and
(A)dS spaces.

The proliferation of \textquotedblleft exotic\textquotedblright\ mixed
symmetry fields poses the question of identifying different representations
that can describe the same spin, possibly in different phases with respect to
a weak/strong coupling limit. This is precisely the subject of duality, which
has been profusely studied along the years in many different contexts
\cite{Hull,Bekaert3}. In the massless case, dual formulations of fields with
spin two and higher in arbitrary dimensions have been derived from a first
order parent action \cite{Boulanger3} based upon the Vasiliev action
\cite{Vasiliev2}. In this case, when the original description of the gauge
fields in dimension $D$ is in terms of totally symmetric tensors, dual
theories in terms of mixed symmetry tensors corresponding to Young tableaux
having one column with ($D-3$) boxes plus ($s-1$) columns with one box have
been obtained \cite{Boulanger3}. A discussion of duality for massless spin two
fields in arbitrary dimensions, which is consistent with the Vasiliev
formulation \cite{Vasiliev2}, has been presented in Ref. \cite{West}. An
alternative construction of theories which are dual to linearized gravity in
arbitrary dimensions has been developed in Ref.\cite{HARIKUMAR}, following the
method of the global shift symmetry applied to the tetrad field.

Contrary to the massless case, dual formulations for massive gravity are not
as well explored and still present issues requiring elucidation. The basic
idea of dualizing the massive Fierz-Pauli (FP) action \cite{Fierz-Pauli},
written in terms of the standard symmetric tensor $h_{\alpha\beta}$, is to
find a formulation where the kinetic contribution of FP yields the mass term
contribution of the corresponding dual theory, and vice-versa. For our
notation and conventions see Ref. \cite{NOTATION}. There are many ways, not
necessarily equivalent, to construct dual theories. A convenient tool to
achieve this is through the use of a first order parent action which contains
both fields and which produces the dual theories via the elimination of the
adequate field using its equations of motion \cite{PL}.

Curtright proposed a dual description of the massive FP action based upon the
mixed symmetry tensor $T_{A[BC]}$ satisfying the same identities as the
linearized spin connection of Einstein's theory in arbitrary dimensions
\cite{CURT}. The corresponding kinetic term was constructed by imposing gauge
invariance under general gauge transformations that respect these identities,
which completely fixed the corresponding relative coefficients. The mass term
was chosen to provide the standard energy-momentum relations for massive
fields. In Ref. \cite{CURT1}, Curtright and Freund (CF) tried different parent
actions in four dimensions to obtain the duality transformation between the FP
action and that corresponding to the mixed symmetry tensor, but they were not
able to obtain such a connection. They could only construct parent actions
where the $h_{\alpha\beta}$ field satisfied the FP action, but the mixed
symmetry tensor $T_{\alpha\lbrack\beta\gamma]}$ was associated to an action
which was different from the one dictated by the gauge symmetry requirements
imposed by their construction. Anyway, the impossibility of obtaining a
dualization of massive FP was not conclusively proved, and they remarked the
necessity of a definitive analysis of the subject.

Motivated by such results, a constructive method based on the parent
Lagrangian approach was pursued in Refs. \cite{CMU1,CMU2}, which dispensed
from the gauge invariance requirements of the action dual to FP. The starting
point of the procedure is a second order Lagrangian in four dimensions, which
depends on the fields $\varphi^{a}$ and their derivatives $\partial_{\mu
}\varphi^{a}$. As the first step, a first order Lagrangian is constructed
using a generalization of a procedure presented in Ref. \cite{Lanczos}, by
introducing, via appropriate Lagrange multipliers $L_{a}^{\mu}$, an adequate
number of invertible auxiliary variables $f_{\mu}^{a}=f_{\mu}^{a}(\varphi
^{a},\partial_{\mu}\varphi^{a})$. This intermediate Lagrangian contains the
fields $f_{\mu}^{a}$ only in algebraic form, and thus they can be eliminated
from the corresponding equations of motion. The resulting Lagrangian contains
the derivatives of the original fields $\varphi^{a}$ times the Lagrange
multipliers $L_{a}^{\mu}$, which become auxiliary variables. A point
transformation in the extended configuration space for the auxiliary variables
$L_{a}^{\mu}$, $L_{a}^{\mu}=\epsilon^{\mu\nu\sigma\tau}H_{a\nu\sigma\tau}$,
incorporates the intuitive idea of duality and yields the parent Lagrangian
from which both dual theories can be obtained. The equations of motion for
$H_{a\nu\sigma\tau}$ make these fields explicit functions of $\varphi^{b}$,
$H_{a\nu\sigma\tau}(\varphi^{b})$, and allows to go back to the original
action after they are substituted in the parent Lagrangian. On the other hand
we can also eliminate the fields $\varphi^{a}$ from the parent Lagrangian
using their own equations of motion, and in such a way we obtain a new theory
that only contains the $H_{a\nu\sigma\tau}$. This new Lagrangian is dual to
the original one, and the equivalence is given by the transformations defined
by the equations of motion of the parent Lagrangian. This approach gives a
parent Lagrangian with a minimum number of fields: the original ones and their
duals. The generalization of this approach to higher order Lagrangians as well
as to arbitrary dimensions is straightforward.

Applying this procedure to the massive spin two field $h_{\left\{  \mu
\nu\right\}  }$, we started with the standard Fierz--Pauli Lagrangian and
constructed a family of first order Lagrangians that contain the symmetric
tensor $h_{\left\{  \mu\nu\right\}  }$ and the mixed symmetry tensor
$T_{\alpha\left[  \mu\nu\right]  }$. Using the equations of motion for
$h_{\left\{  \mu\nu\right\}  }$ we can eliminate this last field, in such a
way that we obtain a set of multiparametric families of Lagrangians dual to
massive Fierz-Pauli, where the dynamics is now contained in the $T_{\alpha
\left[  \mu\nu\right]  }$ field. The unique kinetic term of these dual actions
is fixed by the choice of the FP mass term in the parent action, while only
the corresponding mass terms incorporate the free parameters. A practical
approach to obtain the general structure of such parent Lagrangians amounts to
writing the most general combinations of terms quadratic in the dual fields
$h_{\left\{  \mu\nu\right\}  }$ and $T_{\alpha\lbrack\beta\gamma]}$, plus a
combination of all the possible coupling terms which are linear in each of the
dual fields and include one derivative. The arbitrary coefficients are
partially fixed by eliminating $T_{\alpha\lbrack\beta\gamma]}$ from the parent
Lagrangian and demanding the recovery of the Fierz-Pauli action. After a
careful inspection of all dual Lagrangians obtained with this method it
becomes clear that the Curtright Lagrangian is not obtained.

A key to understand this difficulty is given by the first order action
proposed by West \cite{West}. It has the form of a Lagrangian in our
multiparametric family, but with $h_{\mu\nu}$ without a definite symmetry,
instead of the symmetric one chosen in \cite{CMU1,CMU2}. When $T_{\alpha
\left[  \mu\nu\right]  }$ is eliminated from this first order Lagrangian, the
antisymmetric part of $h_{\mu\nu}$ decouples and becomes irrelevant, so that
we obtain the usual massive Fierz-Pauli theory. On the other hand, when we
eliminate the $h_{\mu\nu}$ field, the presence of its antisymmetric part
alters the dynamics of $T_{\alpha\left[  \mu\nu\right]  }$, thus extending the
families of dual Lagrangians for the massive spin two field to include the
form proposed by Curtright.

A different approach was followed by Zinoviev\ \cite{Zinoviev} based on a
Stueckelberg-like construction for massive tensor fields in Minkoswki as well
as (Anti) de Sitter spaces. For the spin two case in four dimensions he starts
from a first order parent action incorporating the fields ($\omega_{\mu\left[
\alpha\beta\right]  }$, $F^{\left[  \alpha\beta\right]  }$\ , $\pi^{\alpha}$)
which are dual to ($h_{\mu\alpha}$, $A_{\mu}$, $\varphi$). The massive first
order parent action is constructed from the corresponding free actions for the
massless version of the above mentioned fields, plus additional mass terms
which induce a redefinition of the original gauge transformations for the
massless fields in order to preserve a mass dependent gauge invariance of the
full action. In this sense ($F^{\left[  \alpha\beta\right]  }$, $\pi^{\alpha}%
$) together with ($A_{\mu}$,\ $\varphi$) are the auxiliary Stueckelberg fields
for the resulting spin two massive dual fields $h_{\mu\alpha}$\ and
$\omega_{\mu\left[  \alpha\beta\right]  }$. The construction is presented in
four dimensions and no general prescription for arbitrary dimensions is given,
as it has been done for example in the massless case in Ref. \cite{Boulanger3}%
, except for the statement that the method can be easily generalized in such a case.

The paper is organized as follows. In Section II we demonstrate the
equivalence between the dual Zinoviev action, with an adequate gauge fixing,
and the Curtright-Freund action in four dimensions. We also start from the
Zinoviev parent action (which includes a non-symmetrical $h_{\mu\nu}$) and
show that the elimination of some auxiliary fields together with additional
gauge fixing leads to the first order parent action proposed by West
\cite{West} plus a FP mass term. From this parent action we recover, on one
hand, the Fierz-Pauli formulation in terms of the symmetric part of $h_{\mu
\nu}$ and, on the other, the Curtright-Freund dual theory in terms of the
field $T_{\alpha\lbrack\beta\gamma]}$. This duality, described in Section III,
directly relates the description in terms of a symmetric FP field $h_{AB}$ and
a mixed symmetry tensor $T_{B\left[  A_{1}\dots A_{D-2}\right]  }$, satisfying
the cyclic identity, in arbitrary dimensions $D$. This construction does not
require the use of additional Stueckelberg-like fields. The count of the true
degrees of freedom of the dual field $T_{B\left[  A_{1}\dots A_{D-2}\right]
}$ is also performed in this section. The last section contains a summary and
comments on the work. Finally, we conclude with two appendices. In Appendix A
we include some useful expressions incorporating properties of the
antisymmetrized generalized delta function which have proved useful in the
calculations. Appendix B contains the derivation of the Lagrangian constraints
satisfied by the dual field $T_{B\left[  A_{1}\dots A_{D-2}\right]  }$ that
arise from the corresponding equations of motion and which are required in
Section III to obtain the correct number or propagating degrees of freedom.

\section{The Zinoviev approach and the Curtright-Freund formulation in a four
dimensional Minkowski space}

It is relevant to understand the relation between the Zinoviev approach, based
on a first order parent Lagrangian having well defined gauge symmetries
generated by a set of auxiliary fields \cite{Zinoviev}, and the scheme
proposed in Ref. \cite{CMU1}, based on the most general form for the first
order parent Lagrangian containing only the dual fields. In the\ approach of
Ref. \cite{Zinoviev} a duality transformation between Stueckelberg-like
Lagrangians for massive fields is obtained, while in that of Refs.
\cite{CMU1,CMU2} the duality is directly stated at the level of the fields
$h_{\alpha\beta}\;$and $T^{\rho\left[  \mu\nu\right]  }$ corresponding to
different representations for the massive spin two degrees of freedom. When
comparing with works of Zinoviev  one has to keep in mind that his metric is
$diag(+,-,-,-)$ so that we will need to make the appropriate changes of signs
to translate his results into those corresponding to our choice of the metric.
Let us recall that a consistent way of getting the correct relative signs is
to count the total number of $\eta_{\alpha\beta}$ factors in a given
expression, each of which carries a minus sign. Sometimes we make a global
change of sign in the resulting transformed Lagrangian.

\subsection{The Curtright-Freund action from the Zinoviev dual action}

A closer look at the final result of Ref. \cite{Zinoviev} for the dual action
to FP in terms of the field $\omega^{\mu\left[  \alpha\beta\right]  }%
\;$reveals the notable feature that, after gauge fixing, it is equivalent to
the CF action in terms of the field $T^{\rho\left[  \mu\nu\right]  }$, which
is the Hodge dual of $\omega^{\mu\left[  \alpha\beta\right]  }$. To show this
property in a simple way let us start from Eq. (2.12) of
\ Ref\ \cite{Zinoviev} which we rewrite here in flat space (zero cosmological
constant) and in the metric $(-,+,+,+)$%
\begin{align}
\mathcal{L}_{Z} &  =\frac{1}{2}\left(  R^{\mu\nu}R_{\nu\mu}-\frac{1}{3}%
R^{2}\right)  +\frac{1}{6}\left(  \partial^{\alpha}F_{\left[  \alpha
\beta\right]  }\right)  ^{2}+\frac{m}{\sqrt{2}}\left(  \omega^{\mu\left[
\nu\alpha\right]  }\partial_{\alpha}F_{\left[  \mu\nu\right]  }+\frac{1}%
{3}\omega^{\mu}\partial^{\alpha}F_{\left[  \alpha\mu\right]  }\right)
\nonumber\\
&  +\frac{m^{2}}{2}\left(  \omega^{\mu\left[  \alpha\beta\right]  }%
\omega_{\alpha\left[  \mu\beta\right]  }-\frac{1}{3}\omega^{\mu}\omega_{\mu
}\right)  ,
\end{align}
with
\begin{equation}
\omega^{\mu}=\omega_{\alpha}^{\;\;\left[  \alpha\mu\right]  },\;\;R_{\left[
\mu\nu\right]  \left[  \alpha\beta\right]  }=\partial_{\mu}\omega_{\nu\left[
\alpha\beta\right]  }-\partial_{\nu}\omega_{\mu\left[  \alpha\beta\right]
},\;\;R_{\mu}^{\;\;\nu}=R_{\left[  \mu\alpha\right]  }^{\;\ \ \;\;\left[
\alpha\nu\right]  },\ \;R_{\mu\nu}\neq R_{\nu\mu},\label{DEFZ}%
\end{equation}
which is gauge invariant under the transformations
\begin{align}
\delta\omega_{\mu}^{\;\;\left[  \alpha\beta\right]  } &  =\partial_{\mu}%
\theta^{\left[  \alpha\beta\right]  },\;\;\delta F^{\left[  \mu\nu\right]
}=-m\sqrt{2}\theta^{\left[  \mu\nu\right]  },\\
\delta\omega^{\mu\left[  \alpha\beta\right]  } &  =\eta^{\mu\alpha}\xi^{\beta
}-\eta^{\mu\beta}\xi^{\alpha}.\label{XIGI}%
\end{align}
We can use the gauge freedom associated to $\theta^{\left[  \mu\nu\right]
}\;$to set $\;F^{\left[  \mu\nu\right]  }=0$, which leave us with%
\begin{equation}
\mathcal{L}_{Z}=\frac{1}{2}\left(  R^{\mu\nu}R_{\nu\mu}-\frac{1}{3}%
R^{2}\right)  +\frac{m^{2}}{2}\left(  \omega^{\mu\left[  \alpha\beta\right]
}\omega_{\alpha\left[  \mu\beta\right]  }-\frac{1}{3}\omega^{\mu}\omega_{\mu
}\right)  ,
\end{equation}
which is still invariant under the transformations (\ref{XIGI}). After writing
the kinetic part in terms of $\omega^{\mu\left[  \alpha\beta\right]  }$, the
above Lagrangian reduces to%

\begin{equation}
\mathcal{L}_{Z}=\frac{1}{2}\partial^{\beta}\omega_{\mu\left[  \alpha
\beta\right]  }\partial_{\theta}\omega^{\alpha\left[  \mu\theta\right]
}-\frac{1}{6}\left(  \partial_{\alpha}\omega^{\alpha}\right)  ^{2}+\frac
{m^{2}}{2}\left(  \omega^{\mu\left[  \alpha\beta\right]  }\omega
_{\alpha\left[  \mu\beta\right]  }-\frac{1}{3}\omega^{\mu}\omega_{\mu}\right)
. \label{LAG3}%
\end{equation}
It is convenient to split $\omega^{\mu\left[  \alpha\beta\right]  }$ into a
traceless piece $\bar{\omega}^{\mu\left[  \alpha\beta\right]  }$ and the trace
$\omega^{\beta}$,
\begin{equation}
\omega^{\mu\left[  \alpha\beta\right]  }=\bar{\omega}^{\mu\left[  \alpha
\beta\right]  }+\frac{1}{3}\left(  \eta^{\mu\alpha}\omega^{\beta}-\eta
^{\mu\beta}\omega^{\alpha}\right)  ,\;\;\bar{\omega}_{\alpha}^{\;\left[
\alpha\beta\right]  }=0, \label{omega}%
\end{equation}
which transforms as%
\begin{equation}
\delta\bar{\omega}^{\mu\left[  \alpha\beta\right]  }=0,\;\;\;\delta
\omega^{\beta}=3\xi^{\beta}, \label{GTSF}%
\end{equation}
under the remaining gauge symmetry (\ref{XIGI}). Such symmetry\ allows us to
set
\begin{equation}
\omega^{\beta}=0,
\end{equation}
thus reducing the Lagrangian (\ref{LAG3}) to%
\begin{equation}
\mathcal{L}_{Z}=\frac{1}{2}\partial^{\beta}\bar{\omega}_{\mu\left[
\alpha\beta\right]  }\partial_{\gamma}\bar{\omega}^{\alpha\left[  \mu
\gamma\right]  }+\frac{m^{2}}{2}\bar{\omega}^{\mu\left[  \alpha\beta\right]
}\bar{\omega}_{\alpha\left[  \mu\beta\right]  }. \label{LAG4}%
\end{equation}
To make contact with the CF Lagrangian we introduce now the field
$T^{\rho\left[  \mu\nu\right]  }\;$which is dual to $\bar{\omega}%
^{\beta\left[  \mu\gamma\right]  }$,%
\begin{equation}
\bar{\omega}_{\rho\left[  \alpha\beta\right]  }=\epsilon_{\alpha\beta\mu\nu
}T_{\rho}^{\;\;\left[  \mu\nu\right]  }.
\end{equation}
The first property to remark is that the traceless condition upon$\;\bar
{\omega}_{\rho\left[  \alpha\beta\right]  }$ leads to the cyclic identity of
the dual field%
\begin{equation}
T^{\rho\left[  \mu\nu\right]  }+T^{\nu\left[  \rho\mu\right]  }+T^{\mu\left[
\nu\rho\right]  }=0,
\end{equation}
characteristic of the CF field. In terms of this new variable the Lagrangian
(\ref{LAG4}) becomes%
\begin{align}
\mathcal{L}_{Z}  &  =\partial^{\beta}T^{\alpha\left[  \mu\nu\right]  }%
\partial_{\beta}T_{\alpha\left[  \mu\nu\right]  }-2\partial^{\beta}T^{\nu
}\partial_{\beta}T_{\nu}-2\partial_{\mu}T^{\alpha\left[  \mu\nu\right]
}\partial^{\beta}T_{\alpha\left[  \beta\nu\right]  }-\partial_{\beta}%
T^{\beta\left[  \mu\nu\right]  }\partial^{\alpha}T_{\alpha\left[  \mu
\nu\right]  }\nonumber\\
&  -4T_{\nu}\partial_{\mu}\partial_{\beta}T^{\beta\left[  \nu\mu\right]
}+2\partial_{\nu}T^{\nu}\partial^{\beta}T_{\beta}.-m^{2}\left(  T^{\alpha
\left[  \sigma\tau\right]  }T_{\alpha\left[  \sigma\tau\right]  }-2T^{\sigma
}T_{\sigma}\right)  ,
\end{align}
with $T^{\sigma}=T_{\alpha}^{\;\;\left[  \alpha\sigma\right]  }$.
Incorporating now the field strength
\begin{equation}
F_{\nu\left[  \alpha\beta\gamma\right]  }=\partial_{\alpha}T_{\nu\left[
\beta\gamma\right]  }+\partial_{\beta}T_{\nu\left[  \gamma\alpha\right]
}+\partial_{\gamma}T_{\nu\left[  \alpha\beta\right]  },
\end{equation}
corresponding to $T_{\nu\left[  \beta\gamma\right]  }\;$we see that this
Lagrangian is proportional to that of CF \cite{CURT, CURT1}%
\begin{equation}
L_{Z}=\frac{1}{3}\left(  F_{\nu\left[  \alpha\beta\gamma\right]  }%
F^{\nu\left[  \alpha\beta\gamma\right]  }-3F_{\;\;\left[  \alpha\beta
\gamma\right]  }^{\gamma}F_{\sigma}^{\;\;\left[  \alpha\beta\sigma\right]
}-3m^{2}\left(  T^{\alpha\left[  \sigma\tau\right]  }T_{\alpha\left[
\sigma\tau\right]  }-2T^{\sigma}T_{\sigma}\right)  \right)  . \label{LCURTF}%
\end{equation}
\ This establishes that the Zinoviev dual action, with the given gauge fixing,
is in fact the CF action in four dimensions.

\subsection{The Curtright-Freund action from the Zinoviev parent action}

Using Zinoviev approach we should be able to identify the parent Lagrangian at
the level of the relevant fields $\omega_{\alpha\left[  \beta\gamma\right]
}\;$and $h_{\mu\nu}\;$in order to compare with the approach of Refs.
\cite{CMU1,CMU2} and understand how the CF duality in four dimensions can be
obtained from that approach. Notice that here $h_{\mu\nu}\;$is not a
symmetrical field.

To this end we start from the gauge invariant full parent Lagrangian given by
Eqs. (2.1) and (2.5) of Ref. \cite{Zinoviev}
\begin{align}
\mathcal{L}_{h,A,\phi,\omega,F,\pi} &  =\frac{1}{2}\left(  \omega^{\gamma
}\omega_{\gamma}-\omega^{\beta\left[  \alpha\gamma\right]  }\omega
_{\alpha\left[  \beta\gamma\right]  }\right)  -\left(  \omega_{\tau
}^{\;\left[  \nu\alpha\right]  }+\delta_{\tau}^{\alpha}\omega_{\rho
}^{\;\left[  \rho\nu\right]  }-\delta_{\tau}^{\nu}\omega_{\mu}^{\;\left[
\mu\alpha\right]  }\right)  \partial_{\nu}h_{\alpha}^{\;\tau}\nonumber\\
&  -\frac{1}{4}F^{\left[  \alpha\beta\right]  }F_{\left[  \alpha\beta\right]
}+F^{\left[  \mu\nu\right]  }\partial_{\mu}A_{\nu}-\frac{1}{2}\pi^{\alpha}%
\pi_{\alpha}+\pi^{\mu}\partial_{\mu}\phi+\sqrt{3}m\pi^{\mu}A_{\mu}\nonumber\\
&  -\frac{m}{\sqrt{2}}F^{\left[  \mu\nu\right]  }h_{\mu\nu}-\sqrt{2}%
m\omega^{\mu}A_{\mu}+\sqrt{\frac{3}{2}}m^{2}h\phi-m^{2}\phi^{2}+\frac{m^{2}%
}{2}\left(  h^{\alpha\beta}h_{\beta\alpha}-h^{2}\right)  ,\label{GIZLAG}%
\end{align}
which has the following symmetries%
\begin{align}
\delta h_{\mu\nu} &  =\partial_{\mu}\xi_{\nu}+\kappa_{\left[  \mu\nu\right]
}-\frac{m}{\sqrt{2}}\eta_{\mu\nu}\lambda
,\ \ \ \ \ \ \ \ \ \ \ \ \ \ \ \ \ \ \delta h=\partial_{\mu}\xi^{\mu}%
-2\sqrt{2}m\lambda,\nonumber\\
\delta\omega_{\mu}^{\;\left[  \alpha\beta\right]  } &  =\partial_{\mu}%
\kappa^{\left[  \alpha\beta\right]  }+\frac{m^{2}}{2}\left(  \delta_{\mu
}^{\alpha}\xi^{\beta}-\delta_{\mu}^{\beta}\xi^{\alpha}\right)
,\ \ \ \ \ \ \ \delta\omega_{\mu}^{\;\left[  \mu\beta\right]  }=\partial_{\mu
}\kappa^{\left[  \mu\beta\right]  }+\frac{3}{2}m^{2}\xi^{\beta},\nonumber\\
\delta A_{\mu} &  =\frac{m}{\sqrt{2}}\xi_{\mu}+\partial_{\mu}\lambda
,\ \ \ \ \ \ \ \ \ \ \ \ \ \ \ \ \ \ \ \ \ \ \ \ \ \ \ \ \ \ \ \delta
F^{\left[  \alpha\beta\right]  }=-m\sqrt{2}\kappa^{\left[  \alpha\beta\right]
},\nonumber\\
\delta\pi^{\alpha} &  =\sqrt{\frac{3}{2}}m^{2}\xi^{\alpha}%
,\ \ \ \ \ \ \ \ \ \ \ \ \ \ \ \ \ \ \ \ \ \ \ \ \ \ \ \ \ \ \ \ \ \ \ \ \ \delta
\phi=-m\sqrt{3}\lambda.\label{GAUGESYM}%
\end{align}
Again, the corresponding items in Ref. \cite{Zinoviev} are rewritten here in
the metric $(-,+,+,+)$. The basic idea is to eliminate the auxiliary fields
either by their equations of motion or by gauge fixing. The first step is the
elimination of $\pi_{\alpha}$ via its equations of motion, which yield%
\begin{equation}
\pi_{\alpha}=\partial_{\alpha}\phi+\sqrt{3}mA_{\alpha},
\end{equation}
leading to the remaining Lagrangian%
\begin{align}
\mathcal{L}_{h,A,\phi,\omega,F} &  =\frac{1}{2}\left(  \omega^{\gamma}%
\omega_{\gamma}-\omega^{\beta\left[  \alpha\gamma\right]  }\omega
_{\alpha\left[  \beta\gamma\right]  }\right)  -\left(  \omega_{\tau
}^{\;\left[  \nu\alpha\right]  }+\delta_{\tau}^{\alpha}\omega_{\rho
}^{\;\left[  \rho\nu\right]  }-\delta_{\tau}^{\nu}\omega_{\mu}^{\;\left[
\mu\alpha\right]  }\right)  \partial_{\nu}h_{\alpha}^{\;\tau}\nonumber\\
&  -\frac{1}{4}F^{\left[  \alpha\beta\right]  }F_{\left[  \alpha\beta\right]
}+F^{\left[  \mu\nu\right]  }\partial_{\mu}A_{\nu}+\frac{1}{2}\left(
\partial_{\alpha}\phi+\sqrt{3}mA_{\alpha}\right)  ^{2}+\sqrt{\frac{3}{2}}%
m^{2}h\phi-m^{2}\phi^{2}\nonumber\\
&  -\frac{m}{\sqrt{2}}F^{\left[  \mu\nu\right]  }h_{\mu\nu}-\sqrt{2}%
m\omega^{\mu}A_{\mu}+\frac{m^{2}}{2}\left(  h^{\alpha\beta}h_{\beta\alpha
}-h^{2}\right)  .
\end{align}
The above Lagrangian is still invariant under the
transformations\ (\ref{GAUGESYM}), leaving out the transformation $\delta
\pi^{\alpha}$. In this way the gauge freedom can be fixed using the parameters
$\kappa^{\left[  \alpha\beta\right]  }$,\ $\xi_{\mu}$ and $\lambda\;$to set at
zero the fields $F^{\left[  \alpha\beta\right]  }$,\ $A_{\mu}\;$and $\phi
\;$respectively. Then we obtain%
\begin{equation}
\mathcal{L}_{h,;\omega}=\frac{1}{2}\left(  \omega^{\gamma}\omega_{\gamma
}-\omega^{\beta\left[  \alpha\gamma\right]  }\omega_{\alpha\left[  \beta
\gamma\right]  }\right)  -\left(  \omega^{\tau\left[  \nu\alpha\right]  }%
+\eta^{\tau\alpha}\omega^{\nu}-\eta^{\tau\nu}\omega^{\alpha}\right)
\partial_{\nu}h_{\alpha\tau}+\frac{m^{2}}{2}\left(  h^{\alpha\beta}%
h_{\beta\alpha}-h^{2}\right)  .\label{ZREDL1}%
\end{equation}
Observe that we still have a non-symmetrical $h_{\alpha\beta}$ with a specific
choice for the FP mass term. Let us also remark that the Lagrangian
(\ref{ZREDL1}) corresponds to the selection $a=0$ in the parameter of the
corresponding Lagrangian in the Introduction of Ref. \cite{Zinoviev}. This is
exhibited as a simple example of the ambiguities in the dual theory introduced
by constructing the parent Lagrangian with arbitrary coefficients, restricted
only by the condition that after eliminating the field $\omega_{\alpha
\lbrack\beta\gamma]}$ the standard FP theory is recovered, as it is done in
Refs. \cite{CMU1,CMU2}. Nevertheless, given that we arrive at the condition
$a=0$ only after a very particular gauge fixing and field elimination via
equations of motion, we take this as an indication that these two generally
non-commuting and non-unique processes will also introduce ambiguities in the
final Zinoviev Lagrangian containing only the dual propagating field. A more
detailed discussion of this point is given at the end of Section IV.

The above Lagrangian (\ref{ZREDL1}) can be written in terms of the massless
West parent Lagrangian \cite{West}, with well known duality properties. To
this end we introduce the following field transformation%
\begin{equation}
Y^{\tau\left[  \alpha\nu\right]  }=\omega^{\tau\left[  \nu\alpha\right]
}+\eta^{\tau\alpha}\omega^{\nu}-\eta^{\tau\nu}\omega^{\alpha},
\end{equation}
in such a way that%
\begin{equation}
Y^{\alpha}=-2\omega^{\alpha},
\end{equation}
with $Y^{\alpha}=\eta_{\tau\nu}Y^{\tau\left[  \alpha\nu\right]  }$. This
transformation can be inverted as%
\begin{equation}
\omega_{\alpha\left[  \beta\gamma\right]  }=Y_{\alpha\left[  \gamma
\beta\right]  }+\frac{1}{2}\eta_{\alpha\gamma}Y_{\beta}-\frac{1}{2}%
\eta_{\alpha\beta}Y_{\gamma}.
\end{equation}
Applying this transformation to the Lagrangian (\ref{ZREDL1}) we finally
obtain%
\begin{equation}
\mathcal{L}_{h;Y}=\frac{1}{2}\left[  Y^{\tau\left[  \nu\alpha\right]  }\left(
\partial_{\nu}h_{\alpha\tau}-\partial_{\alpha}h_{\nu\tau}\right)
-Y^{\beta\left[  \gamma\alpha\right]  }Y_{\alpha\left[  \gamma\beta\right]
}+\frac{1}{2}Y^{\alpha}Y_{\alpha}+m^{2}\left(  h^{\alpha\beta}h_{\beta\alpha
}-h^{2}\right)  \right]  .\label{w}%
\end{equation}
This is precisely the West action plus a FP type mass term in the notation of
Ref. \cite{Boulanger3}. As shown in this reference, the above Lagrangian in
the massless case leads to the FP one after $Y^{\beta\left[  \alpha
\gamma\right]  }\;$is eliminated using the corresponding equations of
motion.\ The massive case is completely analogous because the equations of
motion for $Y_{\alpha\left[  \beta\gamma\right]  }\;$do not involve the mass
term. Thus, the kinetic energy piece of the action in terms of $h_{\alpha
\beta}$ involves the antisymmetric part $h_{\left[  \alpha\beta\right]  }$
only as a total derivative. The mass term contributes with a term proportional
to $h_{\left[  \alpha\beta\right]  }h^{\left[  \alpha\beta\right]  }$ which
leads to the equation of motion $h_{\left[  \alpha\beta\right]  }=0.\;$It is
rather remarkable that the FP formulation is recovered despite the fact that
$h_{\alpha\beta}$\ is non-symmetrical. The above parent action is not a
particular case of those employed in Refs. \cite{CMU1,CMU2}, where it was
assumed that $h_{\alpha\beta}=h_{\beta\alpha}\;$from the very beginning.\ Let
us recall that the CF case was not obtained in such references.

We will now show, from this point of view, that the dual theory corresponds
precisely to the CF Lagrangian by explicitly eliminating $h_{\alpha\beta}$
from the parent Lagrangian (\ref{w}). The equations of motion for
$h_{\alpha\beta}$ give%

\begin{equation}
h^{\alpha\beta}=\frac{1}{m^{2}}\left(  \partial_{\nu}Y^{\alpha\left[  \nu
\beta\right]  }-\frac{1}{3}\eta^{\beta\alpha}\partial_{\nu}Y^{\nu}\right)
,\qquad h=-\frac{1}{3m^{2}}\partial_{\nu}Y^{\nu}. \label{HFY}%
\end{equation}
After the substitutions (\ref{HFY}) are made in (\ref{w}), the final rescaled
Lagrangian is%

\begin{equation}
\mathcal{\tilde{L}}_{Y}=-2m^{2}\mathcal{L}_{Y}=\partial_{\nu}Y^{\alpha\left[
\nu\beta\right]  }\partial^{\rho}Y_{\beta\left[  \rho\alpha\right]  }-\frac
{1}{3}\left(  \partial_{\nu}Y^{\nu}\right)  ^{2}+m^{2}\left[  Y^{\beta\left[
\gamma\alpha\right]  }Y_{\alpha\left[  \gamma\beta\right]  }-\frac{1}%
{2}Y^{\alpha}Y_{\alpha}\right]  . \label{LZDUALFIN2}%
\end{equation}
In order to make contact with the Lagrangian (\ref{LAG4}) which, as shown in
the previous subsection, leads directly to the CF action we still need to
introduce the traceless field $\bar{\omega}_{\rho\left[  \alpha\beta\right]
}$ according to Eq. (\ref{omega}). In this way the final change of variables
turns out to be%
\begin{equation}
Y^{\tau\left[  \alpha\nu\right]  }=\bar{\omega}^{\tau\left[  \nu\alpha\right]
}+\frac{2}{3}\left(  \eta^{\tau\alpha}\omega^{\nu}-\eta^{\tau\nu}%
\omega^{\alpha}\right)  .
\end{equation}

After substituting in the Lagrangian (\ref{LZDUALFIN2})\ we obtain the dual
one%
\begin{equation}
\mathcal{\tilde{L}}_{\bar{\omega}}=\frac{1}{2}\partial_{\nu}\bar{\omega
}^{\alpha\left[  \beta\nu\right]  }\partial^{\rho}\bar{\omega}_{\beta\left[
\alpha\rho\right]  }+\frac{m^{2}}{2}\left[  \bar{\omega}^{\beta\left[
\alpha\gamma\right]  }\bar{\omega}_{\alpha\left[  \beta\gamma\right]  }%
-\frac{2}{3}\omega^{\alpha}\omega_{\alpha}\right]  . \label{LZDUALFIN3}%
\end{equation}
In fact, the term $\left(  \partial_{\nu}\omega^{\nu}\right)  ^{2}$ cancels
out in the kinetic piece of Lagrangian (\ref{LZDUALFIN2}), while contributions
proportional to $\omega^{\nu}\omega_{\nu}$ in the mass term lead to
$\omega_{\nu}=0$ by the equations of motion. Finally, the Lagrangian
(\ref{LZDUALFIN3}) is identical to\ (\ref{LAG4}), thus leading to the CF final action.

This establishes that the parent Lagrangian of Zinoviev, with the above
specific gauge fixing, gives a duality relation between the FP and the CF
actions for a massive spin two field in four dimensions.

We emphasize that the above duality relation can not be obtained with the
formulation \ presented in Refs. \cite{CMU1,CMU2}. The reason is that there
the tensor $h_{\mu\nu}$ in the parent Lagrangian is taken as symmetric, while
in the parent Lagrangian of Zinoviev it has no definite symmetry. If we
eliminate the field $Y^{\tau\left[  \nu\alpha\right]  }$ both Lagrangians lead
to the FP one with $h_{\mu\nu}$ symmetric, but this difference is crucial when
the eliminated field is $h_{\mu\nu}$, as can easily be visualized as follows.
If we introduce the decomposition
\begin{equation}
h_{\alpha\tau}=h_{\left\{  \alpha\tau\right\}  }+h_{\left[  \alpha\tau\right]
},
\end{equation}
in the Lagrangian (\ref{w}),where $h_{\mu\nu}$ {is not symmetrical}, we get%
\begin{align}
\mathcal{L}_{h;Y} &  =-\tilde{Y}^{\nu\left[  \alpha\tau\right]  }\partial
_{\nu}h_{\left[  \alpha\tau\right]  }-\frac{m^{2}}{2}h^{\left[  \alpha
\tau\right]  }h_{\left[  \alpha\tau\right]  }\nonumber\\
&  +\tilde{Y}^{\nu\left\{  \alpha\tau\right\}  }\partial_{\nu}h_{\left\{
\alpha\tau\right\}  }-\frac{1}{2}\left(  \tilde{Y}^{\nu\left\{  \alpha
\tau\right\}  }\tilde{Y}_{\nu\left\{  \alpha\tau\right\}  }-\tilde{Y}%
^{\nu\left[  \alpha\tau\right]  }\tilde{Y}_{\nu\left[  \alpha\tau\right]
}\right)  +\frac{1}{4}Y^{\alpha}Y_{\alpha}+\frac{m^{2}}{2}\left(  h^{\left\{
\alpha\tau\right\}  }h_{\left\{  \alpha\tau\right\}  }-h^{2}\right)  ,
\end{align}
where%
\begin{equation}
\tilde{Y}^{\nu\left[  \tau\alpha\right]  }=\frac{1}{2}\left(  Y^{\tau\left[
\nu\alpha\right]  }-Y^{\alpha\left[  \nu\tau\right]  }\right)  ,\;\;\;\;\tilde
{Y}^{\nu\left\{  \tau\alpha\right\}  }=\frac{1}{2}\left(  Y^{\tau\left[
\nu\alpha\right]  }+Y^{\alpha\left[  \nu\tau\right]  }\right)
,\;\;\;\;Y^{\alpha}=\eta_{\tau\nu}\tilde{Y}^{\alpha\{\tau\nu\}}.
\end{equation}
Now we can eliminate $h_{\left[  \alpha\tau\right]  }$ using its equation of
motion%
\begin{equation}
h^{\left[  \alpha\tau\right]  }=\frac{1}{m^{2}}\partial_{\nu}\tilde{Y}%
^{\nu\left[  \alpha\tau\right]  },
\end{equation}
obtaining a Lagrangian that only contains $h_{\left\{  \alpha\tau\right\}  }$%
\begin{align}
\mathcal{L}_{h;Y} &  =\frac{1}{2m^{2}}\partial_{\nu}\tilde{Y}^{\nu\left[
\alpha\tau\right]  }\partial^{\mu}\tilde{Y}_{\mu\left[  \alpha\tau\right]
}+\frac{1}{2}\tilde{Y}^{\nu\left[  \alpha\tau\right]  }\tilde{Y}_{\nu\left[
\alpha\tau\right]  }\nonumber\\
&  +\tilde{Y}^{\nu\left\{  \alpha\tau\right\}  }\partial_{\nu}h_{\left\{
\alpha\tau\right\}  }-\frac{1}{2}\tilde{Y}^{\nu\left\{  \alpha\tau\right\}
}\tilde{Y}_{\nu\left\{  \alpha\tau\right\}  }+\frac{1}{4}Y^{\alpha}Y_{\alpha
}+\frac{m^{2}}{2}\left(  h^{\left\{  \alpha\tau\right\}  }h_{\left\{
\alpha\tau\right\}  }-h^{2}\right)  .
\end{align}
If, on the other hand, we consider the Lagrangian (\ref{w}) with
$h_{\alpha\tau}$ purely symmetric its elimination leads to%
\begin{align}
\mathcal{L}_{h;Y} &  =\frac{1}{2}\tilde{Y}^{\nu\left[  \alpha\tau\right]
}\tilde{Y}_{\nu\left[  \alpha\tau\right]  }\nonumber\\
&  +\tilde{Y}^{\nu\left\{  \alpha\tau\right\}  }\partial_{\nu}h_{\left\{
\alpha\tau\right\}  }-\frac{1}{2}\tilde{Y}^{\nu\left\{  \alpha\tau\right\}
}\tilde{Y}_{\nu\left\{  \alpha\tau\right\}  }+\frac{1}{4}Y^{\alpha}Y_{\alpha
}+\frac{m^{2}}{2}\left(  h^{\left\{  \alpha\tau\right\}  }h_{\left\{
\alpha\tau\right\}  }-h^{2}\right)  .
\end{align}
It is clear that in the first case the field $\tilde{Y}^{\nu\left[  \tau
\alpha\right]  }$ is a dynamical one, while in the second one it is null. This
states the difference between the dual theories generated in each case, and
gives us the clue to modify the approach of Refs. \cite{CMU1,CMU2} to generate
a duality transformation that connects the FP theory with a Curtright-type
formulation in arbitrary dimensions.

\section{The parent action and the dual formulation in arbitrary dimensions}

Our aim is the construction of a dual description to the FP formulation for a
massive spin two field $h_{AB}=h_{BA}$ in arbitrary dimensions. We can follow
the procedure developed in Refs. \cite{CMU1,CMU2} to construct first order
parent Lagrangians, but now starting with a nonsymmetric field $h_{AB}$.
According to the discussion there presented, it is possible to construct
several families of dual theories. These parent Lagrangians can be generalized
to arbitrary dimensions. Nevertheless, to be specific, in this work we will
consider only the dual theory generated by a parent Lagrangian that has the
form of the one introduced in Ref. \cite{West} and discussed in Ref.
\cite{Boulanger3}, which corresponds to the Vasiliev description for a
massless spin two, plus the modified FP mass term proposed by Zinoviev. This
is a generalization to unsymmetrical $h_{AB}$ of a special case of the
families just mentioned, and we defer a detailed study of the general
situation in arbitrary dimensions for future work. Thus, in a flat
$D$-dimensional space-time with metric $diag(-+++++,...,+)$ we take%
\begin{align}
S  &  =\frac{1}{2}\int d^{D}x\left[  Y^{C[AB]}\left(  \partial_{A}%
h_{BC}-\partial_{B}h_{AC}\right)  -Y_{C[AB]}Y^{B[AC]}+\frac{1}{(D-2)}%
Y_{\ \ \ [AB]}^{B}Y_{C}^{\ \ \ [AC]}\right. \nonumber\\
&  \left.  +m^{2}\left(  h_{AB}h^{BA}-h^{2}\right)  \right]  ,
\label{PACTDDIM}%
\end{align}
as our parent action. Here the fields are $h_{BC}$ and $Y^{C[AB]}$ , with
$D^{2}$ and $\ D^{2}(D-1)/2$ independent components respectively.\ Redefining
$Y^{C[AB]}\rightarrow-Y^{C[AB]}/\sqrt{2}$\ and $h_{AB}\rightarrow\sqrt
{2}h_{AB}$\ this action becomes the action (4.15) of Ref. \cite{West} plus a
FP mass term, up to a global minus sign.

The derivation of the FP action starting from the action (\ref{PACTDDIM}) is
the same as in Ref. \cite{Boulanger3}, because one needs to solve for
$Y^{C[AB]}$, which does not involve the additional mass term. We only write
the solution in our slightly modified conventions. The resulting expression
for $Y_{B[AC]}$ in terms of $h_{AB}$ is:
\begin{align}
Y_{B[AC]} &  =\frac{1}{2}\left[  \partial_{A}\left(  h_{BC}+h_{CB}\right)
-\partial_{C}\left(  h_{BA}+h_{AB}\right)  -\partial_{B}\left(  h_{AC}%
-h_{CA}\right)  \right]  \nonumber\\
&  +\eta_{BC}\left(  \partial^{D}h_{AD}-\partial_{A}h\right)  -\eta
_{BA}\left(  \partial^{D}h_{CD}-\partial_{C}h\right)  ,\\
Y_{A} &  =Y_{B[AC]}\eta^{\left\{  CB\right\}  }=-\left(  D-2\right)  \left(
\partial_{A}h-\partial^{B}h_{AB}\right)  .\label{YH}%
\end{align}
These expressions allow us to eliminate this field in the action
(\ref{PACTDDIM}). Splitting $h^{AB}$ in its symmetric and antisymmetric parts
\begin{equation}
h^{AB}=h^{\left\{  AB\right\}  }+h^{\left[  AB\right]  },
\end{equation}
and dropping total derivatives we get%
\begin{align}
S &  =\frac{1}{2}\int d^{D}x\left[  -\partial_{A}h_{\left\{  BC\right\}
}\partial^{A}h^{\left\{  CB\right\}  }+2\partial^{B}h_{\left\{  BC\right\}
}\partial_{A}h^{\left\{  AC\right\}  }-2\partial_{A}h\partial_{E}h^{\left\{
AE\right\}  }+\partial_{A}h\partial^{A}h\right.  \nonumber\\
&  \left.  -m^{2}\left(  h_{\left\{  AB\right\}  }h^{\left\{  BA\right\}
}+h_{\left[  AB\right]  }h^{\left[  BA\right]  }-h^{2}\right)  \right]  .
\end{align}
By using the Euler-Lagrange equation of $h_{\left[  AB\right]  }$ we get
$h_{\left[  AB\right]  }=0$, and thus we finally obtain
\begin{align}
S &  =\frac{1}{2}\int d^{D}x\left[  -\partial_{A}h_{\left\{  BC\right\}
}\partial^{A}h^{\left\{  CB\right\}  }+2\partial^{B}h_{\left\{  BC\right\}
}\partial_{A}h^{\left\{  AC\right\}  }-2\partial_{A}h\partial_{E}h^{\left\{
AE\right\}  }+\partial_{A}h\partial^{A}h\right.  \nonumber\\
&  \left.  -m^{2}\left(  h_{\left\{  AB\right\}  }h^{\left\{  BA\right\}
}-h^{2}\right)  \right]  ,
\end{align}
which is precisely the massive FP action in $D$ dimensions. The Euler-Lagrange
equations yield $\left(  D+1\right)  $ constraints, $\partial_{A}h^{\left\{
AB\right\}  }=0$ and $h_{A}^{\ \ A}=0$, and thus the number of degrees of
freedom is
\begin{equation}
\mathcal{F}_{m}^{D}=\frac{D}{2}(D+1)-(D+1)=\frac{D}{2}(D-1)-1.\label{FPDOF}%
\end{equation}

To obtain the dual description we eliminate $h^{AB}$ using its corresponding
equations of motion obtained from the action (\ref{PACTDDIM}), which yield
\begin{equation}
h^{AB}=\frac{1}{m^{2}}\left(  \frac{1}{D-1}\eta^{AB}\partial_{C}Y^{C}%
-\partial_{C}Y^{A[CB]}\right)  , \label{cm4}%
\end{equation}
leading to the following action for $Y_{C[AB]}$
\begin{equation}
m^{2}S=\int d^{D}x\left[  \partial_{A}Y^{C[AB]}\partial^{E}Y_{B[EC]}-\frac
{1}{D-1}(\partial_{A}Y^{A})^{2}+m^{2}\left(  Y_{C[AB]}Y^{B[AC]}-\frac{1}%
{D-2}Y_{A}Y^{A}\right)  \right]  . \label{cm5}%
\end{equation}
To compare with the usual formulation of the Curtright Lagrangian it is useful
to introduce the change of variables
\begin{equation}
Y^{C[AB]}=\bar{w}^{C[AB]}+\frac{1}{(D-1)}(\eta^{CB}Y^{A}-\eta^{CA}Y^{B}),
\label{25tilde}%
\end{equation}
where $\bar{w}^{C[AB]}$ has a null trace, $\bar{w}_{A}^{\ \ [AB]}=0$.
Rescaling the action to absorb the $m^{2}$ factor we obtain%
\begin{equation}
S=\frac{1}{2}\int d^{D}x\left[  \partial_{A}\bar{w}^{C[BA]}\partial^{E}\bar
{w}_{B[CE]}+m^{2}\left(  \bar{w}^{C[AB]}\bar{w}_{B[AC]}-\frac{1}{(D-1)\left(
D-2\right)  }Y^{A}Y_{A}\right)  \right]  ,
\end{equation}
which clearly shows that the trace of $Y^{C[BA]}$ is an irrelevant variable
that can be eliminated from the Lagrangian using its equation of motion. Thus
we finally get%
\begin{equation}
S=\int d^{D}x\frac{1}{2}\left[  \partial_{A}\bar{w}^{C[BA]}\partial^{E}\bar
{w}_{B[CE]}+m^{2}\bar{w}^{C[AB]}\bar{w}_{B[AC]}\right]  . \label{LAGOMEGA}%
\end{equation}
This is the generalization to arbitrary dimensions of the Lagrangian
(\ref{LAG4}).

The derivative term has the gauge symmetries
\begin{equation}
\delta\bar{w}_{C}{}^{[AB]}=\epsilon^{ABM_{1}M_{2}M_{3}...M_{D-2}}%
\partial_{M_{1}}S_{\left\{  {C}M_{2}\right\}  \left[  M_{3}...M_{D-2}\right]
},
\end{equation}%
\begin{equation}
\delta\bar{w}_{C}{}^{[AB]}=\epsilon^{ABM_{1}M_{2}M_{3}\dots M_{D-2}}\left(
\partial_{M_{1}}A_{\left[  CM_{2}M_{3}\dots M_{D-2}\right]  }+\partial
_{C}A_{\left[  M_{1}M_{2}\dots M_{D-2}\right]  }\right)  .
\end{equation}
The mass term breaks these symmetries and assigns to the true degrees of
freedom a mass $m$.

In order to make contact with the usual expression for the Curtright
Langrangian in $D=4$, where the basic field satisfies a cyclic condition, we
need to introduce the Hodge-dual of $\bar{w}^{C\left[  AB\right]  }$%
\begin{equation}
T_{P\left[  Q_{1}Q_{2}...Q_{D-2}\right]  }=\frac{1}{2}\bar{w}_{P}{}^{\left[
AB\right]  }\,\epsilon_{ABQ_{1}Q_{2}...Q_{D-2}}, \label{DUALF}%
\end{equation}
which is a dimension-dependent tensor of rank $\left(  D-1\right)  $
completely antisymmetric in its last $\left(  D-2\right)  $ indices. The
resulting action corresponding to the field $T_{P\left[  Q_{1}Q_{2}%
...Q_{D-2}\right]  }$ will be taken as the dual version of the original FP
formulation. We can invert Eq. (\ref{DUALF}) obtaining%
\begin{equation}
\bar{w}_{C}{}^{\left[  AB\right]  }=-\frac{1}{(D-2)!}T_{C\left[  Q_{1}%
Q_{2}...Q_{D-2}\right]  }\epsilon^{Q_{1}Q_{2}...Q_{D-2}AB}.
\end{equation}
Here we are using the basic definition
\begin{equation}
\epsilon^{A_{1}A_{2}...A_{D-1}A_{D}}\epsilon_{B_{1}B_{2}...B_{D-1}B_{D}%
}=-\delta_{\lbrack B_{1}B_{2}...B_{N-1}B_{N}]}^{[A_{1}A_{2}...A_{N-1}A_{N}]},
\end{equation}
where the required properties of the fully antisymmetrized Kronecker delta
$\delta_{\lbrack B_{1}B_{2}...B_{N-1}B_{N}]}^{[A_{1}A_{2}...A_{N-1}A_{N}]}$,
$N\leq D$, together with its contraction with some relevant tensors, are
written down in the Appendix A. There we have included all the cases relevant
to the calculation and we will not specify the particular relation used in any
of the following steps. The traceless condition upon $\bar{w}_{A\left[
BC\right]  }$ leads to the cyclic identity for the dual field%
\begin{equation}
\epsilon^{Q_{1}Q_{2}...Q_{D-2}AS}T_{S\left[  Q_{1}Q_{2}...Q_{D-2}\right]  }=0.
\label{tcc}%
\end{equation}
It is convenient to introduce the field strength $F^{A\left[  Q_{1}%
Q_{2}...Q_{D-2}Q_{D-1}\right]  }$, which is a tensor of rank $D$, associated
with the potential $T^{A\left[  Q_{1}Q_{2}...Q_{D-2}\right]  }$ given by%
\begin{equation}
F^{A\left[  Q_{1}Q_{2}...Q_{D-2}Q_{D-1}\right]  }=\frac{1}{\left(  D-2\right)
!}\delta_{\lbrack A_{1}A_{2}...A_{D-2}A_{D-1}]}^{[Q_{1}Q_{2}...Q_{D-2}%
Q_{D-1}]}\partial^{A_{1}}T^{A\left[  A_{2}...A_{D-2}A_{D-1}\right]  }.
\end{equation}
In this way $F^{A\left[  Q_{1}Q_{2}...Q_{D-2}Q_{D-1}\right]  }$ satisfies%
\begin{equation}
\epsilon_{CQ_{1}Q_{2}...Q_{D-2}B}\;F^{A\left[  BQ_{1}Q_{2}...Q_{D-2}\right]
}=\left(  D-1\right)  \epsilon_{CQ_{1}Q_{2}...Q_{D-2}B}\partial^{B}T^{A\left[
Q_{1}Q_{2}...Q_{D-2}\right]  }.
\end{equation}
In terms of the Hodge-dual the kinetic part of the Lagrangian becomes%
\begin{align}
\partial_{A}\bar{w}^{C[BA]}\partial^{E}\bar{w}_{B[CE]}  &  =-\frac{1}%
{(D-2)!}\left[  \frac{1}{\left(  D-1\right)  }F_{B}^{\ \ \left[  AQ_{1}%
Q_{2}...Q_{D-2}\right]  }\;F_{\ \ \left[  AQ_{1}Q_{2}...Q_{D-2}\right]  }%
^{B}\right. \nonumber\\
&  \left.  -F_{A}^{\ \ \left[  AQ_{1}Q_{2}...Q_{D-2}\right]  }F_{\ \ \left[
BQ_{1}Q_{2}...Q_{D-2}\right]  }^{B}\right]  ,
\end{align}
while the mass terms acquires the form%
\begin{equation}
\bar{w}^{C[AB]}\bar{w}_{B[AC]}=-\frac{1}{(D-2)!}\left[  T_{B\left[  Q_{1}%
Q_{2}...Q_{D-2}\right]  }T^{B\left[  Q_{1}Q_{2}...Q_{D-2}\right]
}-(D-2)T_{\ \ \left[  CQ_{2}...Q_{D-3}\right]  }^{C}T_{B}^{\ \ \left[
BQ_{2}...Q_{D-3}\right]  }\right]  .
\end{equation}
Thus, the final action dual to FP in arbitrary dimensions can be written%
\begin{align}
S(T)=\int d^{D}x  &  \left\{  -\left[  \frac{1}{\left(  D-1\right)  }%
F_{B}^{\ \ \left[  AQ_{1}..Q_{D-2}\right]  }\;F_{\ \ \left[  AQ_{1}%
..Q_{D-2}\right]  }^{B}-F_{A}^{\ \ \left[  AQ_{1}..Q_{D-2}\right]
}F_{\ \ \left[  BQ_{1}..Q_{D-2}\right]  }^{B}\right]  \right. \nonumber\\
&  \left.  -m^{2}\left[  T_{B\left[  Q_{1}..Q_{D-2}\right]  }T^{B\left[
Q_{1}..Q_{D-2}\right]  }-(D-2)T_{\ \ \left[  CQ_{2}...Q_{D-3}\right]  }%
^{C}T_{B}^{\ \ \left[  BQ_{2}...Q_{D-3}\right]  }\right]  \right\}  ,
\label{SFINAL}%
\end{align}
after an adequate rescaling of the original action. Here the field
$T_{B\left[  Q_{1}Q_{2}...Q_{D-2}\right]  }$ satisfies the cyclic condition
(\ref{tcc}), and the gauge symmetries of the kinetic terms, broken by the mass
term, now become (up to global numerical factors)%
\begin{align}
\delta T_{P\left[  Q_{1}Q_{2}\dots Q_{D-2}\right]  }  &  =\delta_{\lbrack
Q_{1}Q_{2}\dots Q_{D-2}]}^{[M_{1}M_{2}\dots M_{D-2}]}\partial_{M_{1}%
}S_{\left\{  PM_{2}\right\}  \left[  M_{3}M_{4}\dots M_{D-2}\right]  },\\
\delta T_{P\left[  Q_{1}Q_{2}\dots Q_{D-2}\right]  }  &  =\frac{1}%
{(D-2)!}\delta_{\lbrack Q_{1}Q_{2}\dots Q_{D-2}]}^{[M_{1}M_{2}\dots M_{D-2}%
]}\partial_{M_{1}}A_{\left[  PM_{2}\dots M_{D-2}\right]  }+\partial
_{P}A_{\left[  M_{1}M_{2}\dots M_{D-2}\right]  }.
\end{align}
The action (\ref{SFINAL}), which is dual to FP in arbitrary dimensions and
which is free from auxiliary fields, is the main result of this paper. It
reduces to the CF action in four dimensions. We observe that the dual
Lagrangians (\ref{LAG4}) and (\ref{LAGOMEGA}) have identical form when written
in terms of the traceless field $\bar{w}^{C[AB]}$ . Nevertheless this is not
the case after the introduction of the dual field of $\bar{w}^{C[AB]}$ which
will satisfy the cyclic identity.

The action (\ref{SFINAL}) leads to the equation of motion
\begin{align}
&  \left[  \delta_{\lbrack A_{1}A_{2}...A_{D-1}]}^{[AQ_{2}...Q_{D-1}]}%
\delta_{C}^{B}-\left(  D-1\right)  \delta_{\lbrack A_{1}A_{2}..A_{D-1}%
]}^{[BQ_{2}...Q_{D-1}]}\delta_{C}^{A}\right]  \partial^{A_{1}}F_{\ \ \left[
AQ_{2}...Q_{D-1}\right]  }^{C}\nonumber\\
&  -m^{2}\left(  D-2\right)  !\left[  T_{\ \ \ \left[  A_{2}...A_{D-1}\right]
}^{B}-\frac{1}{(D-3)!}\delta_{\lbrack A_{2}.......A_{D-1}]}^{[BM_{3}%
...M_{D-1}]}T_{\ \ \left[  CM_{3}...M_{D-1}\right]  }^{C}\right]  =0,
\end{align}
or more explicitly, in terms of the derivatives of the mixed symmetry tensor
$T_{\ \ \ \left[  A_{2}...A_{D-2}A_{D-1}\right]  }^{B}$
\begin{align}
&  \left[  \left(  D-2\right)  !\delta_{\lbrack A_{1}...A_{D-1}]}%
^{[M_{1}..M_{D-1}]}\delta_{C}^{B}-\delta_{\lbrack A_{1}A_{2}...A_{D-1}%
]}^{[BQ_{2}...Q_{D-1]}}\delta_{\lbrack CQ_{2}..\ Q_{D-1}]}^{[M_{1}...M_{D-1}%
]}\right]  \partial^{A_{1}}\partial_{M_{1}}T_{\ \ \ \left[  M_{2}%
..M_{D-1}\right]  }^{C}\nonumber\\
&  -m^{2}\left[  \left(  D-2\right)  !\right]  ^{2}\left[  T_{\ \ \ \left[
A_{2}...A_{D-1}\right]  }^{B}-\frac{1}{(D-3)!}\delta_{\lbrack A_{2}%
.......A_{D-1}]}^{[BM_{3}...M_{D-1}]}T_{\ \ \left[  CM_{3}...M_{D-1}\right]
}^{C}\right]  =0.\label{tmov}%
\end{align}
In Appendix B we derive the Lagrangian constraints arising from this equation
of motion. The complete set of constraints which the dual field
$T_{\ \ \ \left[  A_{2}...A_{D-1}\right]  }^{B}$ satisfies is
\begin{align}
\epsilon^{Q_{1}Q_{2}...Q_{D-2}AS}T_{S\left[  Q_{1}Q_{2}...Q_{D-2}\right]  } &
=0,\label{c1}\\
T_{\ \ \ \left[  BA_{3}...A_{D-1}\right]  }^{B} &  =0,\label{c2}\\
\partial^{D}T_{\ \ \ \left[  DA_{3}...A_{D-1}\right]  }^{B} &  =0,\label{c3}\\
\partial_{B}T_{\ \ \ \left[  A_{2}A_{3}...A_{D-1}\right]  }^{B} &
=0.\label{c4}%
\end{align}
After implementing these constraints the equation of motion reduces to its
simplest form%
\begin{equation}
\left(  \partial^{2}-m^{2}\right)  T_{\ \ \ \left[  A_{2}...A_{D-1}\right]
}^{B}=0.
\end{equation}
The field $T_{\ \ \ \left[  A_{2}...A_{D-1}\right]  }^{B}$ in $D$ dimensions
has $\mathcal{N}=D^{2}(D-1)/2$ independent components, but it must satisfy the
constraints (\ref{c1}-\ref{c4}). To identify the degrees of freedom it is
convenient to write these constraints in momentum space, and in the rest frame
where $k_{M}=(m,0,0,.......,0,0,0)$. In such a way the constraints (\ref{c3})
and (\ref{c4}) imply that only the components with purely spatial indices are
non null, and give the independent constraints:
\begin{align}
T_{\ \ \ \left[  0I_{3}...I_{D-1}\right]  }^{I_{2}} &  =0\;\;\;\rightarrow
\;\;\;\left(  D-1\right)  \frac{\left(  D-1\right)  !}{2!\left(  D-3\right)
!}\text{ \ constraints,}\\
T_{\ \ \ \left[  I_{2}I_{3}...I_{D-1}\right]  }^{0} &  =0\;\;\;\rightarrow
\;\;\;\frac{\left(  D-1\right)  !}{\left(  D-2\right)  !}\text{
\ constraints,}\\
T_{\ \ \ \left[  0I_{3}...I_{D-1}\right]  }^{0} &  =0\;\;\;\rightarrow
\;\;\;\frac{\left(  D-1\right)  !}{2!\left(  D-3\right)  !}\text{
\ constraints,}%
\end{align}
where now the indices $I_{i}$ run only on spatial values, $I_{i}=1,2,..,D-1$.
Up to this stage we have $\left(  D-1\right)  \left[  D\left(  D-2\right)
+2\right]  /2$ constraints. Taking the above relations into account, the
cyclic identity (\ref{c1}) yields only one additional constraint corresponding
to the choice $A=0$ in the expression%
\begin{equation}
\epsilon^{I_{1}I_{2}...I_{D-2}AI}T_{I\left[  I_{1}I_{2}...I_{D-2}\right]
}=0.\label{cc1}%
\end{equation}
Finally, the constraints (\ref{c2}) lead to
\begin{equation}
T_{\ \ \ \left[  BI_{3}...I_{D-1}\right]  }^{B}=0\ ,\ \ \ \ \ \ B,I_{i}%
=1,...,D-1,\;\;\rightarrow\;\;\frac{(D-1)!}{2(D-3)!}\;\;\text{constraints.}%
\end{equation}
Thus the total number of constraints is%
\begin{equation}
\mathcal{C}=\frac{1}{2}D\left(  D-1\right)  ^{2}+1,
\end{equation}
and the number of degrees of freedom actually is
\begin{equation}
\mathcal{G}=\mathcal{N}-\mathcal{C}=\frac{1}{2}D\left(  D-1\right)  -1,
\end{equation}
which indeed is the same number obtained in Eq. (\ref{FPDOF})\ for
$h_{\left\{  AB\right\}  }$ in the FP formulation.

\section{Final comments}

In this paper we have investigated the possibility of constructing dual
theories for the massive gravitational field in arbitrary dimensions,
following a generalization of the ideas originally proposed in Refs.
\cite{CURT1,CURT}. In these works a dual relation between massive Fierz-Pauli
and a third rank mixed symmetry tensor $T_{A\left[  Q_{1}Q_{2}\right]  }$ was
explored, failing in the attempt of constructing such a relation. The
possibility of using higher rank tensors with mixed symmetry was also
mentioned there, but this approach was not further developed. Thus, the
problem of finding the appropriate parent action providing the duality between
the Fierz-Pauli action and those for the mixed symmetry tensors proposed in
Refs. \cite{CURT1,CURT} has remained an open question. In the present paper we
have shown that such a dual relation can be obtained in four dimensions and we
have also proposed a generalization to arbitrary dimensions in terms of a
$(D-1)$-rank tensor $T_{A\left[  Q_{1}Q_{2}...Q_{D-2}\right]  }$. The
construction can also be presented in terms of the traceless field
$\bar{\omega}_{A\left[  BC\right]  }$, dual to $T_{A\left[  Q_{1}%
Q_{2}...Q_{D-2}\right]  }$, in terms of which the action has the same form in
any dimension.

The motivation for our construction is rooted in the attempt to understand the
relation between the Zinoviev approach \cite{Zinoviev}, based on a first order
parent Lagrangian having well defined gauge symmetries generated by a set of
auxiliary fields, and the approach proposed in Ref. \cite{CMU1}, based on the
most general form for a first order parent Lagrangian containing only the dual
fields. In the Zinoviev formalism a duality transformation between
Stueckelberg-like Lagrangians for massive fields is obtained, while in that of
Refs. \cite{CMU1,CMU2} the duality is directly stated at the level of the
fields corresponding to different representations for the massive spin two
degrees of freedom.

With the purpose of making contact between the two approaches, in Section II
we take as the starting point the first order parent Lagrangian (2.1) plus the
terms (2.5) of Ref. \cite{Zinoviev}, in the flat space limit, which depends on
the fields $\omega_{\mu}^{\ \ {\left[  \alpha\beta\right]  }}$, $h_{\mu
}^{\ \ \alpha}$, $F^{\left[  \alpha\beta\right]  }$, $A_{\mu}$, $\pi^{\alpha}%
$, and $\phi$. After eliminating $\pi^{\alpha}$ and being consistent with the
remaining gauge symmetries, we use a gauge fixing such that all the auxiliary
fields become null, and only the spin two dual fields $\omega_{\mu
}^{\ \ \left[  \alpha\beta\right]  }$and $h_{\mu}^{\ \ \alpha}$ remain. From
here, implementing an adequate transformation, we show that this gauge fixed
parent Lagrangian is precisely equivalent to that proposed by West
\cite{West}, plus a FP type mass term, in the notation of Ref.
\cite{Boulanger3}. This parent Lagrangian leads to massive Fierz-Pauli after
eliminating $Y^{\beta\left[  \alpha\gamma\right]  }$. On the other hand, after
eliminating $h_{\alpha\beta}$, we have shown that it is equivalent to the
Curtright-Freund action in four dimensions. This establishes that in four
dimensions the parent Lagrangian of Zinoviev with the above specific gauge
fixing is equivalent to the West parent Lagrangian, which provides a duality
relation between the Fierz-Pauli and the Curtright-Freund actions for a
massive spin two field. We emphasize that the above duality relation between
Fierz-Pauli and Curtright-Freund was not obtained in Refs. \cite{CMU1,CMU2}.
The reason is very simple: in such references the tensor $h_{\mu\nu}$ in the
parent Lagrangian is taken as symmetric, while in the parent Lagrangian
introduced by West it has no definite symmetry.

On the basis of the last observation, the formalism of Refs. \cite{CMU1,CMU2}
has been extended to arbitrary dimensions in Section III, by replacing the
symmetric $h_{\left\{  AB\right\}  }$ tensor in the particular parent
Lagrangian (\ref{PACTDDIM}) by one without a definite symmetry. In such a way
we obtain a new description for the massive Fierz-Pauli gravitation in terms
of a mixed symmetry tensor $T_{S\left[  Q_{1}Q_{2}...Q_{D-2}\right]  }$, based
on an action whose kinetic term satisfies the gauge symmetries compatible with
the cyclic condition (\ref{tcc}). We have also identified the propagating
modes of this theory, showing that they correspond to purely transversal
components of a traceless field. Within the parent Lagrangian formalism there
are additional possibilities, starting from the general structure for the
first order Lagrangian discussed in Ref. \cite{CMU1} together with a
nonsymmetric $h_{AB}$, which are not discussed here.

A comment regarding the two approaches considered in this work is now in
order. Our parent Lagrangian construction is based on the most general first
order Lagrangian that contains only a given field and its dual, provided that
the elimination of the dual field yields the adequate theory for the original
field. This most general Lagrangian may depend on several parameters, and thus
the elimination of the original field leads to a multiparametric family of
dual Lagrangians, i.e. for a given theory we can in general construct several
dual descriptions. On the other hand, the Zinoviev approach is based on a
different perspective, which leads to the construction of a Stueckelberg-type
parent Lagrangian that contains the original and the dual variables together
with a set of auxiliary fields required to implement certain gauge symmetries.
To derive the dual Lagrangian in terms of the corresponding propagating
physical field it is necessary not only to choose some necessary gauge
fixings, but also to use some equations of motion. This can be readily
appreciated in the four dimensional parent Lagrangian (2.1) and (2.5) of Ref.
\cite{Zinoviev}, which starts with 55 independent fields plus 11 arbitrary
functions to be gauge fixed. Going from the remaining 44 variables to the
final 10 degrees of freedom requires either some field eliminations via
equations of motion or some field redefinitions that unify certain
combinations. Clearly this adds a lot of freedom to the final result. In this
way, different gauge fixings will lead to dual Lagrangians which are
equivalent from the point of view of belonging to the same gauge orbits of the
original Lagrangian, but not necessarily equivalent among themselves, in the
sense that they cannot be connected by modifying the actions with boundary
terms. The alternative gauge fixed Lagrangians lead to different patterns for
eliminating the remaining auxiliary variables by using their equations of
motion. This opens up additional possibilities for the appearance of further
non equivalent dual Lagrangians. The very different starting points of both
approaches makes it very difficult, if at all possible, to establish a general
relation between them. In this paper we have only shown that the Zinoviev
approach with a given gauge fixing leads to a dual Lagrangian also contained
in the first order parent Lagrangian approach. We defer for further work the
study of the possible general connections between these two approaches.

In a nutshell we can summarize our results by saying that the parent
Lagrangian looked for by Curtright and Freund for a massive spin two field in
arbitrary dimensions is simply given by the Lagrangian of West \cite{West}
completed by the Fierz-Pauli mass term arising from the Zinoviev approach
\cite{Zinoviev}, and involving a non-symmetrical rank-two tensor.

\appendix

\section{Properties of the generalized antisymmetric Kronecker delta}

We summarize some relations including the antisymmetrized generalized delta
function together with its contractions with various antisymmetric tensors.

The completely antisymmetrized generalized delta function $\delta_{[M_{1}%
M_{2}...M_{N}]}^{[A_{1}A_{2}....A_{N}]}$ in $D$ dimensions having $N!$ terms
($N\leq D$), is defined as
\begin{equation}
\delta_{[M_{1}M_{2}...M_{N}]}^{[A_{1}A_{2}....A_{N}]}=\det\left[
\begin{array}
[c]{cccc}%
\delta_{M_{1}}^{A_{1}} & \delta_{M_{1}}^{A_{2}} & ... & \delta_{M_{1}}^{A_{N}%
}\\
\delta_{M_{2}}^{A_{1}} & \delta_{M_{2}}^{A_{2}} & ... & \delta_{M_{2}}^{A_{N}%
}\\
... & ... & ... & ...\\
\delta_{M_{N}}^{A_{1}} & \delta_{M_{N}}^{A_{2}} & ... & \delta_{M_{N}}^{A_{N}}%
\end{array}
\right]  , \label{A1}%
\end{equation}
having the basic decomposition property%

\begin{equation}
\delta_{[M_{1}...M_{N}]}^{[A_{1}....A_{N}]}=\sum_{I=1}^{N}\left(  -1\right)
^{\left(  I-1\right)  }\delta_{M_{I}}^{A_{1}}\delta_{[M_{1}...M_{I-1}%
\,\;M_{I+1}...M_{N}]}^{[A_{2}...A_{I-1}A_{I}A_{I+1}....A_{N}]}. \label{A2}%
\end{equation}
One important property is the contraction of the first $I$\ indices
\begin{equation}
\delta_{[A_{1}...A_{I}M_{I+1}..M_{N}]}^{[A_{1}..A_{I}A_{I+1}....A_{N}]}%
=\frac{\left(  D-N+I\right)  !}{(D-N)!}\delta_{[M_{I+1}...M_{D}]}%
^{[A_{I+1}....A_{D}]}. \label{A3}%
\end{equation}

The following contractions follow directly from the definition%
\begin{align}
\delta_{[M_{1}...M_{N}]}^{[A_{1}....A_{N}]}S^{\left[  M_{1}...M_{N}\right]
}T_{\left[  A_{1}...A_{N}\right]  }  &  =N!\ S^{\left[  M_{1}...M_{N}\right]
}T_{\left[  M_{1}...M_{N}\right]  },\\
S^{M\left[  M_{1}\dots M_{N-1}\right]  }T_{A\left[  A_{1}\dots A_{N-1}\right]
}\;\delta_{[MM_{1}\dots M_{N-1}]}^{[AA_{1}\dots A_{N-1}]}  &  =(N-1)!\ \left[
S^{Q\left[  M_{1}\dots M_{N-1}\right]  }T_{Q\left[  M_{1}\dots M_{N-1}\right]
}\right. \nonumber\\
&  \left.  -\left(  N-1\right)  \ S^{P\left[  QM_{1}\dots M_{N-2}\right]
}T_{Q\left[  PM_{1}\dots M_{N-2}\right]  }\right]  ,\label{A6}\\
S^{B}{}_{{\left[  R_{1}\dots R_{N-1}\right]  }}\;T_{A}{}^{\left[
Q_{1}...Q_{N-1}\right]  }\ \delta_{[BQ_{1}\dots Q_{N-1}]}^{[AR_{1}\dots
R_{N-1}]}  &  =(N-1)!\;\left[  S_{A}{}^{\left[  Q_{1}...Q_{N-1}\right]
}\;T^{A}{}_{\left[  Q_{1}\dots Q_{N-1}\right]  }\right. \nonumber\\
&  \left.  -(N-1)\ S_{A}{}^{\left[  AQ_{1}\dots Q_{N-2}\right]  }\;T^{B}%
{}_{\left[  BQ_{1}\dots Q_{N-2}\right]  }\right]  .
\end{align}

\section{The Lagrangian constraints on $T_{A\left[  Q_{1}Q_{2}\dots
Q_{D-2}\right]  }$}

Starting from the equations of motion (\ref{tmov}) we derive the constraints
(\ref{c2}), (\ref{c3}) and (\ref{c4}), which together with the cyclic identity
(\ref{c1}) provide the correct number of degrees of freedom for the dual field
$T_{\ \ \ \left[  DA_{3}...A_{D-1}\right]  }^{B}$.

Contracting a derivative with one of the antisymmetric indices in the equation
of motion (\ref{tmov}) we obtain the following first set of constraints%
\begin{equation}
\partial^{D}T^{B}{}_{\left[  DA_{3}...A_{D-1}\right]  }=\frac{1}{(D-3)!}%
\delta_{\lbrack DA_{3}\dots A_{D-1}]}^{[BM_{3}...M_{D-1}]}\partial^{D}T^{C}%
{}_{\left[  CM_{3}\dots M_{D-1}\right]  }.\label{B1}%
\end{equation}
Contracting next one of the antisymmetric free indices in (\ref{tmov}) , for
example $A_{2}$, with $B$ we get
\begin{align}
&  \left(  D-2\right)  !\delta_{\lbrack BA_{1}A_{3}\dots A_{D-1}]}%
^{[M_{1}M_{2}\dots M_{D-1}]}\partial^{A_{1}}\partial_{M_{1}}T_{\ \ \ \left[
M_{2}M_{3}\dots M_{D-1}\right]  }^{B}-\delta_{\lbrack BA_{1}A_{3}\dots
A_{D-1}]}^{[BQ_{2}Q_{3}\dots Q_{D-1}]}\delta_{\lbrack CQ_{2}\dots Q_{D-1}%
]}^{[M_{1}M_{2}\dots M_{D-1}]}\partial^{A_{1}}\partial_{M_{1}}T_{\ \ \ \left[
M_{2}M_{3}\dots M_{D-1}\right]  }^{C}\nonumber\\
&  +m^{2}\left[  \left(  D-2\right)  !\right]  ^{2}\left[  T_{\ \ \ \left[
BA_{3}\dots A_{D-1}\right]  }^{B}-\frac{1}{(D-3)!}\delta_{\lbrack BA_{3}\dots
A_{D-1}]}^{[BM_{3}\dots M_{D-1}]}T_{\ \ \left[  CM_{3}\dots M_{D-1}\right]
}^{C}\right]  =0.\label{B2}%
\end{align}
In fact the second term with derivatives is proportional to the first one.
This can be proved using an adequate expansion of the antisymmetric delta
according to (\ref{A2}). In this way the first derivative term can be written
\begin{equation}
\delta_{\lbrack BA_{1}A_{3}\dots A_{D-1}]}^{[M_{1}M_{2}\dots M_{D-1}]}%
\partial^{A_{1}}\partial_{M_{1}}T_{\ \ \ \left[  M_{2}\dots M_{D-1}\right]
}^{B}=\delta_{\lbrack A_{1}A_{3}\dots A_{D-1}]}^{[M_{2}M_{3}\dots M_{D-1}%
]}\left(  \partial^{A_{1}}\partial_{B}T_{\ \ \ \left[  M_{2}M_{3}\dots
M_{D-1}\right]  }^{B}-(D-2)\partial^{A_{1}}\partial_{M_{2}}T_{\ \ \ \left[
B\dots M_{D-1}\right]  }^{B}\right)  \label{B3}%
\end{equation}
Using the relations (\ref{A2}) and (\ref{A3}),\ the second derivative term
yields
\begin{align}
&  \delta_{\lbrack BA_{1}A_{3}\dots A_{D-1}]}^{[BQ_{2}Q_{3}\dots Q_{D-1}%
]}\delta_{\lbrack CQ_{2}\dots\ \ \ Q_{D-1}]}^{[M_{1}M_{2}\dots M_{D-1}%
]}\partial^{A_{1}}\partial_{M_{1}}T_{\ \ \ \left[  M_{2}\dots M_{D-1}\right]
}^{C}=\nonumber\\
&  =\left(  D-2\right)  !\delta_{\lbrack BA_{1}A_{3}\dots A_{D-1}]}%
^{[BQ_{2}Q_{3}\dots Q_{D-1}]}\left(  \partial^{A_{1}}\partial_{C}%
T_{\ \ \ \left[  Q_{2}Q_{3}\dots Q_{D-1}\right]  }^{C}-(D-2)\partial^{A_{1}%
}\partial_{Q_{2}}T_{\ \ \ \left[  CQ_{3}\dots Q_{D-1}\right]  }^{C}\right)
\nonumber\\
&  =2\left(  D-2\right)  !\delta_{\lbrack BA_{1}A_{3}\dots A_{D-1]}}%
^{[M_{1}M_{2}\ \ \dots M_{D-1}]}\partial^{A_{1}}\partial_{M_{1}}%
T_{\ \ \ \left[  M_{2}\dots M_{D-1}\right]  }^{B}.\label{B4}%
\end{align}
and Eq. (\ref{B2}) becomes%
\begin{equation}
\delta_{\lbrack BA_{1}\dots\;\;A_{D-1}]}^{[M_{1}M_{2}\dots M_{D-1}]}%
\partial^{A_{1}}\partial_{M_{1}}T_{\ \ \ \left[  M_{2}\dots M_{D-1}\right]
}^{B}+2\left(  D-2\right)  !m^{2}\left[  T_{\ \ \left[  CM_{3}\dots
M_{D-1}\right]  }^{C}\right]  =0.\label{B5}%
\end{equation}
Here we have used Eq. (\ref{A3}) in order to rewrite the square bracket
proportional to $m^{2}$ in Eq.(\ref{B2}).\ The derivative term in Eq.
(\ref{B5}) can also be written%
\begin{equation}
\delta_{\lbrack BA_{1}\dots A_{D-1}]}^{[M_{1}...M_{D-1}]}\partial^{A_{1}%
}\partial_{M_{1}}T_{\ \ \left[  M_{2}\dots M_{D-1}\right]  }^{B}=(D-2)\left[
\left(  D-3\right)  !\partial^{A_{1}}\partial_{B}T_{\ \ \left[  A_{1}%
A_{3}\dots A_{D-1}\right]  }^{B}-\delta_{\lbrack A_{1}A_{3}\dots A_{D-1}%
]}^{[M_{2}M_{3}\dots M_{D-1}]}\partial^{A_{1}}\partial_{M_{2}}T_{\ \ \left[
BM_{3}\dots M_{D-1}\right]  }^{B}\right]
\end{equation}
and using the first set of constraints already obtained, (\ref{B1}), we have
\begin{equation}
\delta_{\lbrack A_{1}A_{3}\dots A_{D-1}]}^{[M_{2}\dots M_{D-1}]}%
\partial^{A_{1}}\partial_{M_{2}}T_{\ \ \ \left[  B...M_{D-1}\right]  }%
^{B}=(D-3)!\partial_{B}\partial^{D}T_{\ \ \ \left[  DA_{3}...A_{D-1}\right]
}^{B},\label{B6}%
\end{equation}
which finally yields%
\begin{equation}
\delta_{\lbrack BA_{1}A_{3}\dots A_{D-1}]}^{[M_{1}M_{2}\dots M_{D-1}]}%
\partial^{A_{1}}\partial_{M_{1}}T_{\ \ \ \left[  M_{2}\dots M_{D-1}\right]
}^{B}=0.\label{B7}%
\end{equation}
In such a way Eq. (\ref{B5}) reduces to a second set of constraints%
\begin{equation}
T_{\ \ \ \left[  BA_{3}\dots A_{D-1}\right]  }^{B}=0.\label{B8}%
\end{equation}

Combining the constraints (\ref{B1}) and (\ref{B8}), together with (\ref{c1})
we have the following set of constraints for $T_{S\left[  Q_{1}Q_{2}\dots
Q_{D-2}\right]  }$
\begin{align}
T_{\ \ \ \left[  BA_{3}\dots A_{D-1}\right]  }^{B}  &  =0,\\
\partial^{D}T_{\ \ \ \left[  DA_{3}...A_{D-1}\right]  }^{B}  &  =0.
\end{align}
The last set of constraints%
\begin{equation}
\partial_{B}T_{\ \ \ \left[  A_{2}A_{3}\dots A_{D-1}\right]  }^{B}=0
\end{equation}
is obtained by explicitly rewriting the cyclic identity (\ref{c1}) and
subsequently contracting a derivative with the unsymmetrized index in the
first term. This contraction will appear among the antisymmetric indices in
the remaining terms of the sum, each of which will be identically zero in
virtue of the constraints (\ref{c3}).

\section*{Acknowledgements}

We would like to thank the referee for valuable observations and suggestions.
LFU would like to thank useful discussions with J. A. Garc\'{\i}a. A. K.
acknowledges institutional support from CDCHT-ULA under project C-1506-07-05-B
and the Program High Energy Physics Latinamerican-European Network (HELEN).
R.M. acknowledges partial support from CONICET-Argentina. L.F.U is partially
supported by projects CONACYT \# 55310 and DGAPA-UNAM-IN109108. R.M. and
L.F.U. have been partially supported by a project of international cooperation CONACYT-CONICET.


\begin{thebibliography}{99}
\bibitem {NOTATION}The indices of $D$-dimensional tensors will be denoted with
capital latin letters, while greek letters will be specifically used for the
$4$-dimensional ones. Square brackets will indicate antisymmetrization and
curly brackets symmetrization. The Minkowski metric is denoted by $\eta
_{AB}=\eta_{BA}=diag(-1,+1,\dots+1)$ and the completely antismmetric tensor by
$\epsilon_{A_{1}A_{2}\dots A_{D}}$. Only in these cases the symmetry is not
explicitly stated by the corresponding brackets.
\end{thebibliography}
\end{document}